# Real-time volumetric image reconstruction and 3D tumor localization based on a single x-ray projection image for lung cancer radiotherapy

**Ruijiang Li, Xun Jia, John H. Lewis, Xuejun Gu, Michael Folkerts, Chunhua Men, and Steve B. Jiang[a]**

*Department of Radiation Oncology, University of California San Diego, La Jolla, CA 92037-0843*

## Abstract

**Purpose**: To develop an algorithm for real-time volumetric image reconstruction and 3D tumor localization based on a single x-ray projection image for lung cancer radiotherapy.

**Methods**: Given a set of volumetric images of a patient at *N* breathing phases as the training data, we perform deformable image registration between a reference phase and the other *N-1* phases, resulting in *N-1* deformation vector fields (DVFs). These DVFs can be represented efficiently by a few eigenvectors and coefficients obtained from principal component analysis (PCA). By varying the PCA coefficients, we can generate new DVFs, which, when applied on the reference image, lead to new volumetric images. We then can reconstruct a volumetric image from a single projection image by optimizing the PCA coefficients such that its computed projection matches the measured one. The 3D location of the tumor can be derived by applying the inverted DVF on its position in the reference image. Our algorithm was implemented on graphics processing units (GPUs) to achieve real-time efficiency. We generated the training data using a realistic and dynamic mathematical phantom with 10 breathing phases. The testing data were 360 cone beam projections corresponding to one gantry rotation, simulated using the same phantom with a 50% increase in breathing amplitude.

**Results**: The average relative image intensity error of the reconstructed volumetric images is 6.9% ± 2.4%. The average 3D tumor localization error is 0.8 mm ± 0.5 mm. On an NVIDIA Tesla C1060 GPU card, the average computation time for reconstructing a volumetric image from each projection is 0.24 seconds (range: 0.17 and 0.35 seconds).

**Conclusions**: We have shown the feasibility of reconstructing volumetric images and localizing tumor positions in 3D in near real-time from a single x-ray image.

## 1. Introduction

Management of tumor motion is a challenging and important problem for conformal lung cancer radiotherapy. Poorly managed tumor motion can lead to poor target coverage and an unnecessarily high dose to normal tissues[1]. Therefore, precise knowledge of real-time lung tumor motion during the treatment delivery is essential for the effectiveness of lung cancer radiotherapy[2-8]. Many treatment machines are equipped with an on-board imaging system, which usually consists of one kilo-voltage x-ray source and one imager. However, it is very difficult to localize lung tumors in 3D space based on a single x-ray projection image, especially without the aid of implanted radio-opaque markers. Zeng *et al.* estimated 3D respiratory motion from cone beam projections based on a generic B-spline motion model[9]. However, because of the large number of parameters in the model, many projections over a 180° rotation have to be used. The estimation is a retrospective and lengthy process (computation takes several hours on MATLAB) and cannot be used for real-time tumor localization. Therefore, a more efficient lung motion model is necessary for such purposes. Zhang *et al.* developed a lung motion model based on principal component analysis (PCA), which can efficiently represent the lung motion with only a few eigenvectors and coefficients[10]. The PCA lung motion model was recently shown to bear a close relationship with the physiological 5D lung motion model proposed by Low *et al.*[11] on a theoretical basis[12]. What's more, the implicit regularization of the PCA model allows one to derive dynamic lung motion with very limited information, and yet in a reasonably accurate way.

In this work, we describe a new algorithm based on the PCA lung motion model and demonstrate its ability of reconstructing volumetric images and extracting 3D tumor motion information in real time from a single x-ray projection in a non-invasive (no marker implantation required), accurate, and efficient way.

## 2. Methods

Given a set of volumetric images of a patient at $N$ breathing phases as the training data, we perform deformable image registration between a reference phase and the other $N$-$1$ phases, resulting in a set of $N$-$1$ deformation vector fields (DVFs). This set of DVFs can be represented efficiently by a few eigenvectors and coefficients obtained from PCA. By varying the PCA coefficients, we can generate new DVFs, which, when applied on the reference image, lead to new volumetric images. We then can reconstruct a volumetric image from a single projection image by optimizing the PCA coefficients such that its computed projection image matches the measured one. The 3D location of the tumor can be derived by applying the inverted DVF on its position in the reference image. In reality, the set of training volumetric images can either be 4DCT available from treatment simulation or 4DCBCT available during patient setup.

### 2.1 PCA lung motion model

In the PCA model, the DVF relative to a reference image as a function of space and time is approximated by a linear combination of the sample mean vector and a few eigenvectors corresponding to the largest eigenvalues, *i.e.*,

$$\mathbf{x}(t) \approx \bar{\mathbf{x}} + \sum_{k=1}^{K} \mathbf{u}_k w_k(t) . \quad (1)$$

where $\mathbf{u}_k$ are the eigenvectors obtained from PCA and are functions of space only. The scalars $w_k(t)$ are PCA coefficients and are functions of time only. It is worth mentioning that the eigenvectors are fixed after PCA and it is the evolution of the PCA coefficients that drives the new volumetric image and the dynamic lung motion in real time.

There are primarily two reasons why the PCA lung motion model is suitable for this work. First, PCA provides the best linear representation of the data in the least-mean-square sense (efficiency). Second, the PCA motion model imposes inherent regularization on its representation. One can show that if two voxels move similarly, their motion represented by PCA will also be similar[12]. The combined effect is that a few scalar variables (PCA coefficients) are sufficient to dynamically deform the lung in a reasonably accurate way.

### 2.2 Image reconstruction using PCA model

After we have obtained a parameterized PCA lung motion model, we seek a set of optimal PCA coefficients such that the projection of the corresponding volumetric image matches with the measured x-ray projection. However, the computed and measured projection images may have different intensity levels. Here we assume there exists a linear relationship between them. The cost function is:

$$\begin{aligned} &\min . J(\mathbf{w}, a, b) = \left\| \mathbf{P} \cdot \mathbf{f}(\mathbf{x}, \mathbf{f}_0) - a \cdot \mathbf{y} - b \cdot \mathbf{1} \right\|_2^2 \\ &s.t. \quad \mathbf{x} = \overline{\mathbf{x}} + \mathbf{U} \cdot \mathbf{w} \end{aligned} \tag{2}$$

where, $\mathbf{U}$ is matrix whose columns are the PCA eigenvectors and $\mathbf{w}$ is a vector comprised of the PCA coefficients to be optimized, $\mathbf{x}$ is the parameterized DVF, $\mathbf{f}_0$ is the reference image, $\mathbf{f}$ is the reconstructed image, $\mathbf{y}$ is the measured projection image, and $\mathbf{P}$ is a projection matrix which computes the projection image of $\mathbf{f}$. For clarity of notation, we have suppressed the time index under $\mathbf{w}$, $\mathbf{x}$, $\mathbf{y}$ and $\mathbf{P}$.

To find the optimal values for $\mathbf{w}, a, b$, the algorithm alternates between the following 2 steps:

$$\text{step 1: } \mathbf{w}_{n+1} = \mathbf{w}_n - \mu_n \cdot \frac{\partial J}{\partial \mathbf{w}_n} \tag{3}$$

$$\text{step 2: } (a_{n+1}, b_{n+1})^T = \left(\mathbf{Y}^T \mathbf{Y}\right)^{-1} \mathbf{Y}^T \mathbf{P} \mathbf{f}_{n+1} \tag{4}$$

where, $\mathbf{Y} = [\mathbf{y}, \mathbf{1}]$, and $\dfrac{\partial J}{\partial \mathbf{w}} = \dfrac{\partial \mathbf{x}}{\partial \mathbf{w}} \cdot \dfrac{\partial \mathbf{f}}{\partial \mathbf{x}} \cdot \dfrac{\partial J}{\partial \mathbf{f}} = 2 \cdot \mathbf{U}^T \cdot \dfrac{\partial \mathbf{f}}{\partial \mathbf{x}} \cdot \mathbf{P}^T \cdot \left(\mathbf{P} \cdot \mathbf{f} - a \cdot \mathbf{y} - b \cdot \mathbf{1}\right)$.

Given the updated PCA coefficients and hence DVF in step 1, the reconstructed image $\mathbf{f}_{n+1}$ is found through trilinear interpolation. Accordingly, $\partial \mathbf{f} / \partial \mathbf{x}$ has to be consistent with the interpolation process in order to get the correct gradient. It turns out that $\partial \mathbf{f} / \partial \mathbf{x}$ is a linear combination of the spatial gradients of the reference image evaluated at the neighboring eight grid points, weighted by the appropriate fractional part of the DVF. Step 1 is a gradient descent method with variable $\mathbf{w}$ and fixed *a*, *b*, where the step size $\mu_n$ is found by Armijo's rule for line search. In step 2, the update for *a*, *b* is the unique minimizer of the cost function with fixed $\mathbf{w}$. Therefore, the cost function always decreases at each step. Note that the cost function is lower bounded by zero. The above alternating algorithm is guaranteed to converge for all practical purposes. The algorithm stops whenever the norm of the gradient is sufficiently small or the maximum number of iterations is reached. At this point, we have reconstructed the volumetric image which is the best estimate of the current patient geometry in the least mean square sense, given the reference image, the parameterized DVF, and the measured projection image.

### 2.3 Tumor localization via deformation inversion

In order to get the current tumor position, it is important to distinguish between two different kinds of DVFs: push-forward DVF and pull-back DVF. The DVF found by Eq. (2) is a pull-back

DVF. It cannot be used directly to calculate the new tumor position. To do that, we need its inverse, *i.e.*, the push-forward DVF. Here, we adopt an efficient fixed-point algorithm for deformation inversion[13] and calculate tumor position.

### 2.4 GPU implementation

Recently, general-purpose computing on graphic processing units (GPUs) has offered superb efficiency for computationally intensive tasks in medical imaging and therapy applications[14, 15]. To achieve real-time efficiency, we have implemented our algorithm using compute unified device architecture (CUDA) as the programming environment.

## 3. Materials

The algorithm was tested using a non-uniform rational B-spline (NURBS) based cardiac-torso (NCAT) phantom[16]. This mathematical phantom is based on data from the Visible Human Project, and is very flexible, maintaining a high level of anatomical realism (*e.g.*, a beating heart, detailed bronchial trees, *etc.*). The respiratory motion was developed based on basic knowledge of respiratory mechanics. We generated a dynamic NCAT phantom composed of 10 breathing phases as the training data, with a 3D tumor motion magnitude of 1.6 cm and a breathing period of 4 seconds. The dimension of each volumetric image is: 256×256×120 (voxel size: 2×2×2.5 $mm^3$). We used the end of exhale (EOE) phase as the reference image and performed deformable image registration (DIR) between the EOE phase and all other phases. The DIR algorithm used here is a fast demons algorithm implemented on GPU[15]. Then PCA was performed on the nine DVFs from DIR and three PCA coefficients and eigenvectors were kept in the PCA lung motion model.

For testing purposes, we try to reconstruct a volumetric image and derive the 3D tumor location from each of 360 cone beam projections, which are generated from the NCAT phantom and uniformly distributed over one full gantry rotation. The phantom has a 4 s breathing period and a 50% increase in breathing amplitude (3D tumor motion magnitude: 2.4 cm) relative to the training data. The gantry rotation lasts one minute, resulting in 15 breathing cycles and 24 projections per cycle. Corresponding to 24 projections, we have 24 volumetric images at 24 breathing phases. These 24 volumetric images will be used as ground truth test images to evaluate the accuracy of the reconstructed images. The imager has a physical size of 40×30 $cm^2$. For efficiency considerations, we down-sample every measured projection image to a resolution of 200×150 (pixel size: 2×2 $mm^2$).

To quantify the accuracy of our reconstruction algorithm, we use the relative image error defined as: $e=\sqrt{\sum_i\left(f_i-f_i^*\right)^2 \Big/ \sum_i f_i^{*2}}$ , where $f$ is the reconstructed volumetric image, $f^*$ is the ground truth image from the NCAT phantom, and the summation is taken over all the voxel indices. We quantify the localization accuracy by the 3D root-mean-square (RMS) error.

The accuracy of our algorithm is limited by the accuracy of training DVFs derived with the demons algorithm. The best accuracy we can achieve is the accuracy of the DIR between the ground truth test image and the reference image using demons. We therefore deform the reference image to the ground truth test image using the same demons algorithm and compute the relative image error of the deformed image against the ground truth test image. This error, termed as *deformation error*, in contrast to the *reconstruction error* of the reconstructed image, is used as the benchmark to evaluate our algorithm. Similarly, we also compute the 3D RMS deformation error.

## 4. Results

Figure 1 shows the "measured" projection image, simulated from the NCAT phantom at the end of inhale phase, so that it is maximally different from the reference image. We performed 10 iterations and further iterations were found to have little influence on the results. The relative image reconstruction error is initially 35% and approaches to the deformation error (10.9% compared with 8.3%) after 9 iterations. Figure 2 shows the sagittal and coronal views of the absolute difference images between the reference image and three other images, namely the ground truth test image, the image reconstructed by our algorithm, and the deformed reference image using demons. The 3D RMS tumor localization error is about 0.9 mm. (anterior-posterior: 0.2 mm; lateral: 0.8 mm; superior-inferior: 0.4 mm).

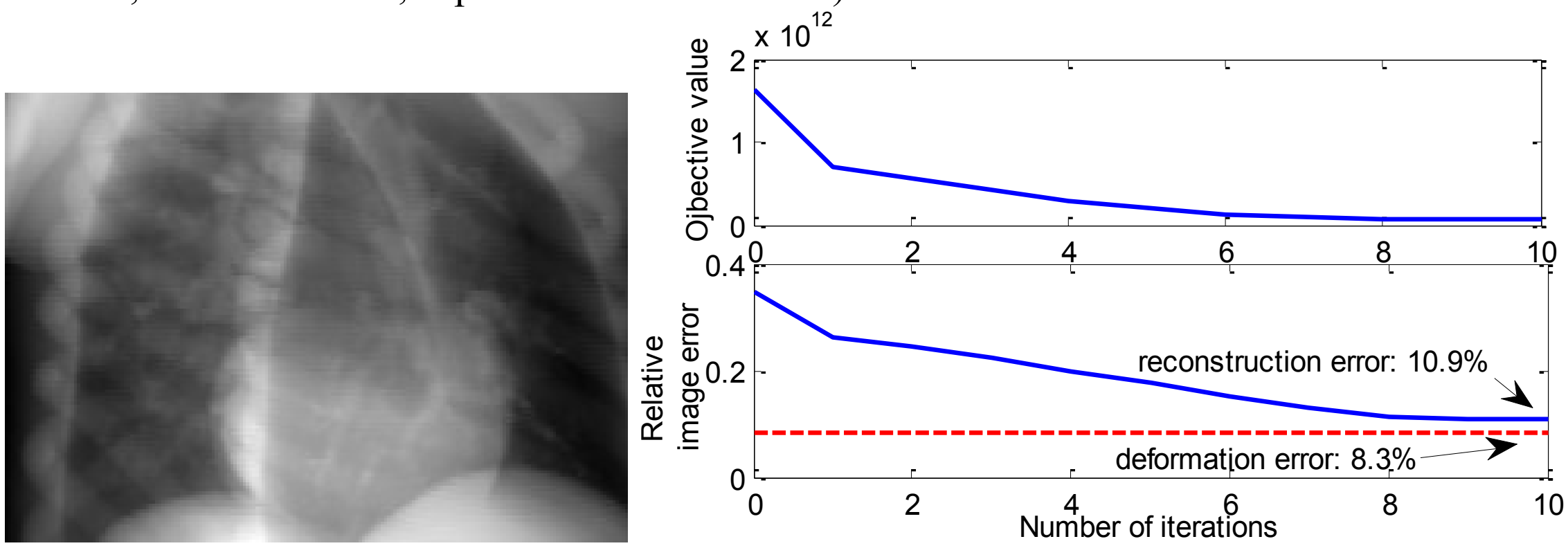

**Fig. 1.** Left: "measured" projection of the test image at a right posterior oblique (RPO) angle; Right: objective value and the relative image reconstruction error at each iteration.

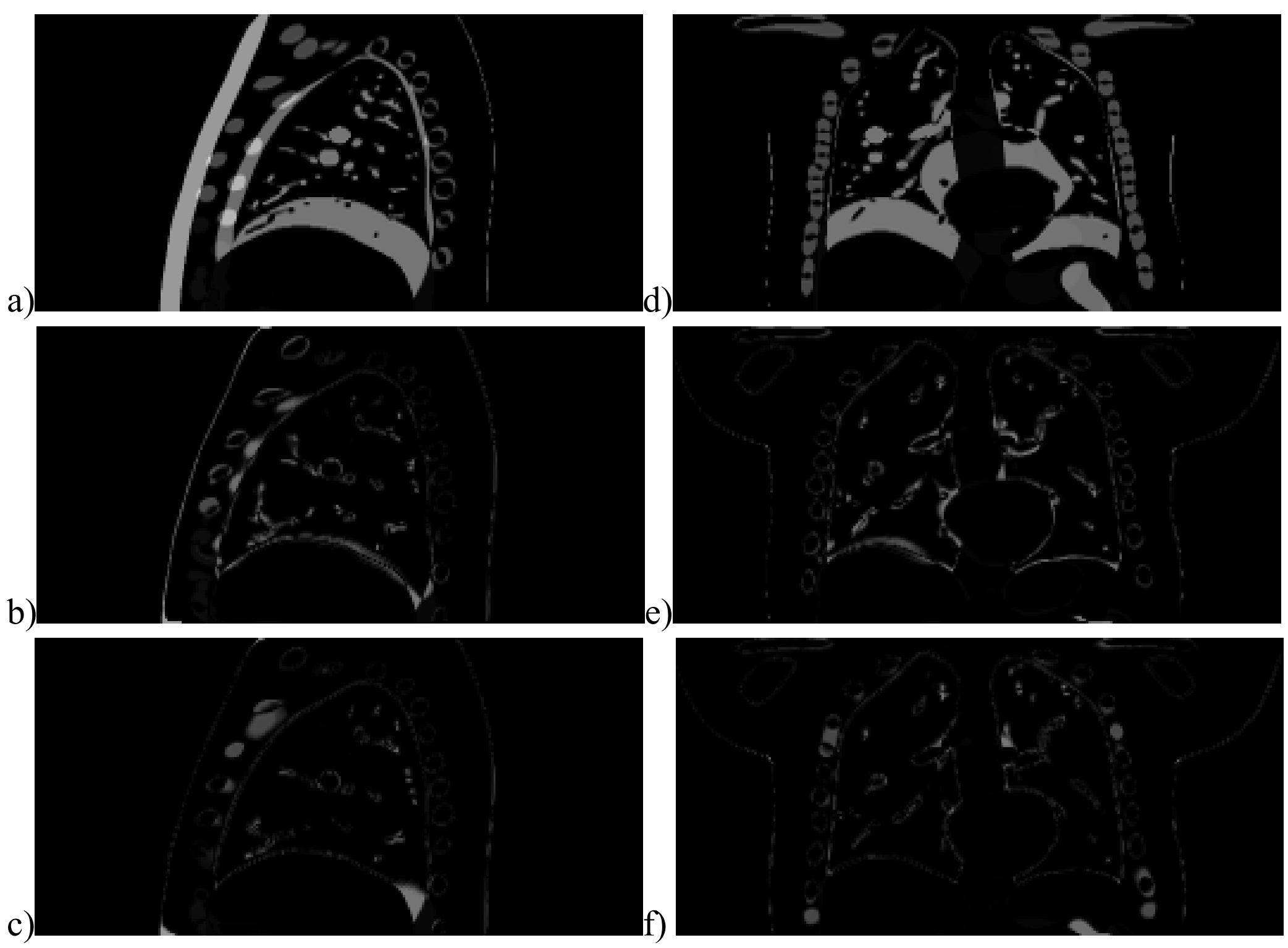

**Fig. 2.** Left column: sagittal view of the absolute difference image between: a) ground truth test and reference images, b) ground truth test image and image reconstructed using a single projection, c) ground truth test image and the deformed reference image using demons. Tumor is a round object near the center of the slice. Right column: same as left column, except for coronal view. Tumor is a round object in the right lung.

Overall, for the 360 cone beam projections, the average relative image reconstruction error is 6.9% ± 2.4%. The average 3D tumor localization error is 0.8 mm ± 0.5 mm and is not affected by projection angles (see Fig. 3). In comparison, the average relative image deformation error is 5.4% ± 2.2% and 3D RMS tumor localization error from deformation is 0.2 mm ± 0.1 mm.

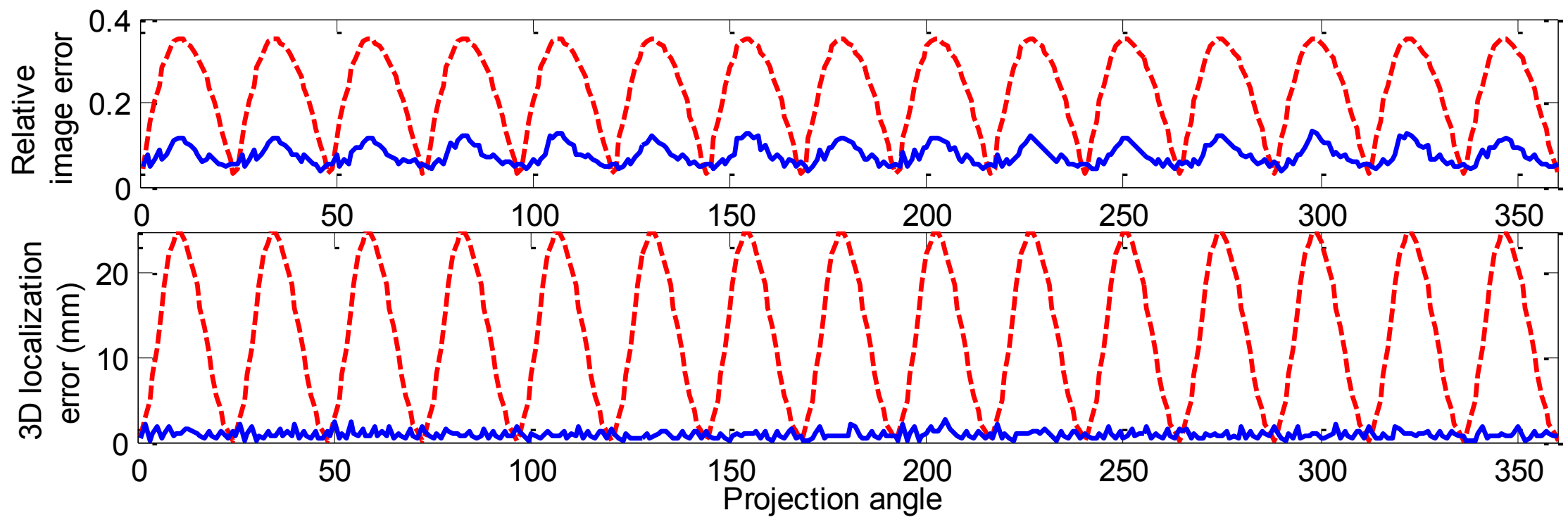

**Fig. 3.** Top row: relative image error between the ground truth test image and: reference image (*red*); image reconstructed using the proposed algorithm (*blue*) as a function of cone beam projection angle. Bottom row: same as top row, except for 3D localization error.

We initialized the PCA coefficients to those at its previous frame considering the high-frequency image acquisition (6 Hz) relative to breathing. The image reconstruction and tumor localization for each projection is achieved in less than 0.4 seconds using an NVIDIA Tesla C1060 GPU card. Particularly, depending on the difference between the test image and reference image, the algorithm converges between 0.17 seconds and 0.35 seconds (average 0.24 seconds). Compared with an implementation on MATLAB 7.7 running on a PC with a Quad 2.67 GHz CPU, which takes around 15 minutes to converge, our GPU version achieves a speedup factor of more than 2500.

## 5. Discussions and Conclusion

We have shown that it is feasible to reconstruct volumetric images and localize 3D tumor positions from a single x-ray projection in a non-invasive, accurate, and efficient way. The relative image reconstruction error is only marginally larger than its lower bound defined by the deformation error. The average 3D RMS tumor localization error is below 1 mm. By utilizing the massive computing power of GPUs, we were able to reconstruct a volumetric image and derive 3D tumor locations from one projection within 0.4 seconds.

To investigate if the projection image resolution of 200×150 is sufficient, we tested the algorithm on a higher resolution of 1000×750 (pixel size: 0.4×0.4 $mm^2$). The improvement of using the fine-resolution projection images is negligible (both image error and localization error are the same as above). In fact, further reduction in imager resolution does not have noticeable effects on the results unless it drops below 40×30, for the phantom case tested here. This demonstrates the ability of the PCA model to infer lung motion using limited information. For real clinical cases, higher resolution may be needed which will be investigated in our future work.

We plan to test the accuracy of our algorithm on clinical data. There are several complicating factors that may affect the accuracy of the current algorithm. First, the x-ray energies used to generate the training volumetric images (such as 4DCT) and cone beam projection images may be different. Second, there will be scattering and other physical effects that may degrade the quality of the cone beam projection images in patient data. In both cases, a linear relationship between the image intensity of the computed and measured projection images may not be accurate. Some preprocessing (*e.g.*, a nonlinear transform) may be needed. We will investigate the application of more general similarity measures such as mutual information. This work has focused on x-ray projections with rotational geometry. The same principle can be easily applied to those with fixed-angle geometry, such as fluoroscopy. The only difference is that the projection matrix will be constant for fixed-angle geometry instead of varying at each angle for the rotational geometry considered in this work. To further speed up the computation, we can also use the predicted PCA coefficients from previous histories to serve as the initial guess for optimization. This procedure might be important when the imaging frequency is low.

**Acknowledgment**

This work is supported in part by Varian Master Research Agreement. We would like to thank Dr. Paul Segars for providing the source code to generate NCAT phantom and NVIDIA for providing GPU cards for this project.

**References**

[1]P.J. Keall, G.S. Mageras, J.M. Balter, R.S. Emery, K.M. Forster, S.B. Jiang, J.M. Kapatoes, D.A. Low, M.J. Murphy, B.R. Murray, C.R. Ramsey, M.B. Van Herk, S.S. Vedam, J.W. Wong, and E. Yorke, "The management of respiratory motion in radiation oncology report of AAPM Task Group 76," Med Phys. **33**(10): 3874-900 (2006).

[2]L.I. Cervino, A.K. Chao, A. Sandhu, and S.B. Jiang, "The diaphragm as an anatomic surrogate for lung tumor motion," Phys Med Biol. **54**(11): 3529-3541 (2009).

[3]R.I. Berbeco, S. Nishioka, H. Shirato, and S.B. Jiang, "Residual motion of lung tumors in end-of-inhale respiratory gated radiotherapy based on external surrogates," Med Phys. **33**(11): 4149-56 (2006).

[4]J.D. Hoisak, K.E. Sixel, R. Tirona, P.C. Cheung, and J.P. Pignol, "Correlation of lung tumor motion with external surrogate indicators of respiration," Int J Radiat Oncol Biol Phys. **60**(4): 1298-306 (2004).

[5]H. Wu, Q. Zhao, R.I. Berbeco, S. Nishioka, H. Shirato, and S.B. Jiang, "Gating based on internal/external signals with dynamic correlation updates," Phys Med Biol. **53**(24): 7137-7150 (2008).

[6]Y. Cui, J.G. Dy, G.C. Sharp, B. Alexander, and S.B. Jiang, "Multiple template-based fluoroscopic tracking of lung tumor mass without implanted fiducial markers," Phys Med Biol. **52**(20): 6229-42 (2007).

[7]Q. Xu, R.J. Hamilton, R.A. Schowengerdt, B. Alexander, and S.B. Jiang, "Lung Tumor Tracking in Fluoroscopic Video Based on Optical Flow," Medical Physics. **35**(12): 5351-5359 (2008).

[8]T. Lin, L.I. Cervino, X. Tang, N. Vasconcelos, and S.B. Jiang, "Fluoroscopic tumor tracking for image-guided lung cancer radiotherapy," Phys Med Biol. **54**(4): 981-992 (2009).

[9]R. Zeng, J.A. Fessler, and J.M. Balter, "Estimating 3-D respiratory motion from orbiting views by tomographic image registration," IEEE Trans Med Imaging. **26**(2): 153-63 (2007).

[10]Q. Zhang, A. Pevsner, A. Hertanto, Y.C. Hu, K.E. Rosenzweig, C.C. Ling, and G.S. Mageras, "A patient-specific respiratory model of anatomical motion for radiation treatment planning," Med Phys. **34**(12): 4772-81 (2007).

[11]D.A. Low, P.J. Parikh, W. Lu, J.F. Dempsey, S.H. Wahab, J.P. Hubenschmidt, M.M. Nystrom, M. Handoko, and J.D. Bradley, "Novel breathing motion model for radiotherapy," Int J Radiat Oncol Biol Phys. **63**(3): 921-9 (2005).

[12]R. Li, J.H. Lewis, X. Jia, T. Zhao, J. Lamb, D. Yang, D.A. Low, and S.B. Jiang. "PCA-based lung motion model." in *16th International Conference on the Use of Computers in Radiation Therapy. accepted*. 2010. Amsterdam, Netherlands.

[13]M. Chen, W. Lu, Q. Chen, K.J. Ruchala, and G.H. Olivera, "A simple fixed-point approach to invert a deformation field," Med Phys. **35**(1): 81-8 (2008).

[14]F. Xu and K. Mueller, "Real-time 3D computed tomographic reconstruction using commodity graphics hardware," Physics in Medicine and Biology. **52**(12): 3405-3419 (2007).

[15]X. Gu, H. Pan, Y. Liang, R. Castillo, D. Yang, D. Choi, E. Castillo, A. Majumdar, T. Guerrero, and S.B. Jiang, "Implementation and evaluation of various demons deformable image registration algorithms on a GPU," Phys Med Biol. **55**(1): 207-19 (2010).

[16]W.P. Segars, D.S. Lalush, and B.M.W. Tsui, "Modeling respiratory mechanics in the MCAT and spline-based MCAT phantoms," IEEE Transactions on Nuclear Science. **48**(1): 89-97 (2001).

[a] Email: sbjiang@ucsd.edu